\documentclass[conference]{IEEEtran}
\usepackage{soul}
\usepackage{amsmath,amssymb,amsfonts}
\usepackage{algorithmic}
\usepackage{graphicx}
\usepackage{textcomp}
\usepackage{xcolor}
\def\BibTeX{{\rm B\kern-.05em{\sc i\kern-.025em b}\kern-.08em
    T\kern-.1667em\lower.7ex\hbox{E}\kern-.125emX}}
\usepackage{graphicx}
\usepackage{longtable}
\usepackage{wrapfig}
\usepackage{rotating}
\usepackage[normalem]{ulem}
\usepackage{amsmath}
\usepackage{amssymb}
\usepackage{capt-of}
\usepackage{hyperref}
\usepackage{comment}
\usepackage{tikz}
\usepackage{pgfplots}
\pgfplotsset{compat=1.18}
\usepackage{pgfplotstable}
\usepackage[style=ieee, maxbibnames=99, backend=bibtex]{biblatex}
\usepackage{xcolor}
\usepackage{xspace}
\usepackage[caption=false,font=normalsize,labelfont=sf,textfont=sf]{subfig}
\usepackage{todonotes}

\colorlet{samcolor}{blue}

\newcommand{\mcmcts}{(MC)\textsuperscript{2}TS\xspace}

\usetikzlibrary{shapes.misc,arrows,arrows.meta}
\tikzset{selected/.style={ultra thick,>=latex}}

\title{Shackling Uncertainty using Mixed Criticality \\ in Monte-Carlo Tree Search}

\author{
  \IEEEauthorblockN{Franco Cordeiro, Samuel Tardieu, Laurent Pautet}
  \IEEEauthorblockA{LTCI, Télécom Paris, Institut Polytechnique de Paris, Paris, France}
  \IEEEauthorblockA{\{%
    \href{mailto:franco.cordeiro@telecom-paris.fr}{franco.cordeiro},
    \href{mailto:samuel.tardieu@telecom-paris.fr}{samuel.tardieu}, \href{mailto:laurent.pautet@telecom-paris.fr}{laurent.pautet}\}@telecom-paris.fr
  }
}

\newcommand{\HI}{\textrm{\textsc{hi}}\xspace}
\newcommand{\LO}{\textrm{\textsc{lo}}\xspace}

\begin{document}
\maketitle

\begin{abstract}
  
  In the world of embedded systems, optimizing actions with the uncertain costs of multiple resources in order to achieve an objective is a complex challenge. Existing methods include plan building based on Monte Carlo Tree Search (MCTS), an approach that thrives in multiple online planning scenarios. However, these methods often overlook uncertainty in worst-case cost estimations.
  A system can fail to operate/function before achieving a critical objective when actual costs exceed optimistic worst-case estimates, even if replanning is considered. Conversely, a system based on pessimistic worst-case estimates would lead to resource over-provisioning even for less critical objectives. To solve similar issues, the Mixed Criticality (MC) approach has been developed in the real-time systems community. In this paper, we propose to extend the MCTS-based heuristic in three directions.

  Firstly, we reformulate the concept of MC to account for uncertain worst-case costs, including optimistic and pessimistic worst-case estimates. High-criticality tasks must be executed regardless of their uncertain costs. Low-criticality tasks are either executed in low-criticality mode utilizing resources up-to their optimistic worst-case estimates, or executed in high-criticality mode by degrading them, or discarded when resources are scarce. In such cases, resources previously devoted to low-criticality tasks are reallocated to high-criticality tasks.

  Secondly, although the MC approach was originally developed for real-time systems, focusing primarily on worst-case execution time as the only uncertain resource, our approach extends the concept of resources to deal with several resources at once, such as the time and energy required to perform an action.
  
  Finally, we propose an extension of MCTS with MC concepts, which we refer to as \mcmcts{}, to efficiently adjust resource allocation to uncertain costs according to the criticality of actions. We demonstrate our approach in an active perception scenario.
 Our evaluation shows \mcmcts{} outperforms the traditional MCTS regardless of whether the worst case estimates are optimistic or pessimistic.
  
 \end{abstract}


\section{Introduction}
\label{sec:Intro}
The challenges of autonomous robot mission planning are multifaceted, particularly in scenarios of \textit{active perception} where a robot actively collects information about an area. Several geographically distributed sensors can collect data without being connected to a network. A drone collects data from each sensor via Bluetooth. It can be critical to complete such an objective before the sensor runs out of battery or memory. On its way, the drone may want to take photos. In such missions, drones are assigned multiple objectives of different criticality.
The difficulties are amplified by uncertainties in real-world situations, where factors like the energy required to move under unpredictable environmental conditions (\textit{e.g.,} strong wind, heavy rain) significantly impact mission planning and execution \cite{Haruna2023_Survey}. Depending on these conditions, the drone may want to give priority to critical objectives and forego less critical ones. Moreover, energy is not the only ressource required to keep track of, as the robot must reach an objective before the sensor runs out of memory.

\begin{figure}[ht]
  \centering
  \includegraphics[width=\linewidth]{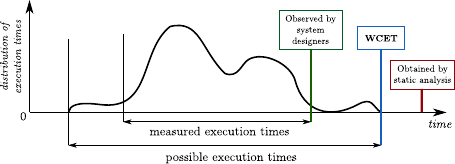}
 \caption{Optimistic and pessimistic WCET estimates} 
  \label{fig:wcet}
\end{figure}

The problem of cost uncertainty has been studied by the real-time community, particularly when dealing with uncertainty in \textit{worst-case execution time} (WCET) estimates \cite{Baruah2018}. Indeed, the WCET evaluation often cannot be carried out with precision (see \figurename~\ref{fig:wcet}). The system designers may obtain optimistic WCET through measurement-based methods while certification authorities may require pessimistic WCET obtained through static code analysis techniques \cite{wilhelm2008worst}. However, the WCET can be bounded by either low or high estimates depending on safety guarantees required by the function. Relying solely on high WCET estimations in system design may result in unnecessary oversizing. Conversely, leaning towards low WCET estimations can lead to scenarios where execution budget constraints are exceeded before task completion.

This challenge led to the emergence of the concept of \textit{Mixed Criticality} (MC) in real-time systems \cite{Burns2017}. In this paradigm, tasks are systematically classified according to their criticality level, \textit{i.e.} by assessing the consequences of a task failure. High critical tasks are usually imperative to the survival of the system, while low critical tasks are related to services. In case of a resource shortage, the lower criticality tasks are degraded in order to guarantee the execution of higher criticality ones. This resource reallocation ensures failure probabilities fall within an acceptable predetermined range.

In context of action planning, a variety of heuristics have been proposed to navigate the complexities of decision-making \cite{Haruna2023_Survey}. Amid these heuristics, \textit{Monte Carlo Tree Search} (MCTS) emerges as a popular choice for online planning in which an agent needs to navigate through a tree-like search space to find an optimal path or make optimal decisions.

In this context, MCTS is not inherently designed to balance guaranteeing execution of some objectives and maximizing overall objective completion. As it is based on the Monte-Carlo method, its decision is strongly influenced by the model's most probable scenarios. Adapting to optimistic scenarios can lead to a resource shortage, while adapting to pessimistic scenarios can lead to oversizing the system.

To address these issues, our contribution is threefold. Firstly, we reformulate the MC approach in the context of uncertainty in the costs. Secondly, although the MC approach was originally designed for real-time systems, focusing on worst-case execution time as the only uncertain resource, we extend the approach by generalizing the notion of resources, in particular to energy. Finally, we design \mcmcts{}, which is an extension of MCTS with MC consideration. It aims to anticipate changes in resource costs and adjust the decision-making process to optimize objective completion, resource utilization and system reliability in terms of criticality compliance.

The rest of the paper is organized as follows. In section \ref{sec:Model}, we describe the system model and define some notations. Then we formulate the problem formally in Section \ref{sec:Problem}. We describe our \mcmcts{} heuristic in Section \ref{sec:Approach}. In Section \ref{sec:Eval}, we evaluate our heuristic and demonstrate the improvement in performance brought by our approach.

\section{System Model}
\label{sec:Model} 

\newcommand{\xbs}[1]{\boldsymbol{x^{\textrm{\tiny #1}}}}
\newcommand{\obs}[1]{\boldsymbol{o^{\textrm{\tiny #1}}}}

We consider a robot which decides to follow a sequence of upcoming actions $\boldsymbol{x}$ = $(x_1,x_2,..., x_n)$ to fulfill objectives that can be measured by an objective function $g$. The system designer assigns a \textit{criticality level} to an action based on the consequences of its failure. In this work, we only consider two criticality levels, meaning actions are either part of the set of high-criticality actions $\boldsymbol{x}^\HI$ or the set of low-criticality actions $\boldsymbol{x}^\LO$. Both sets must be non-empty for our approach to have any positive impact on the execution of $\boldsymbol{x}$.

The system runs in a \textit{criticality mode} that can be either high (\HI-mode) or low (\LO-mode). It starts in \LO-mode and when any resource consumption exceeds its allocated budget in \LO-mode, a mode switch to \HI-mode occurs. A high-criticality action (\HI-action) must be carried out whether the system is running in \LO-mode or \HI-mode. A low-criticality action (\LO-action) is executed only when the system is in \LO-mode. This means \HI-actions must not depend on the execution of \LO-actions. However, if the system switches to \HI-mode while the \LO-action is ongoing, this action will continue until completion. This difference with the classical MC model will be explained in section \ref{sec:Approach}. If the system switches to \HI-mode, future \LO-actions are discarded in order to free up resources for the \HI-actions currently in the plan. A seamless transition between modes is required: execution must continue running smoothly as far as \HI-actions are concerned. The system may later switch back to \LO-mode if all resource consumption return to normal.

Every action $x_i$ is associated with an actual cost $c_i$, a tuple whose elements represent the different resources we track. In our case, the cost tuple would be $c_i = [\textrm{d}_i, \textrm{e}_i]$. $c_i[\textrm{duration}]$ (resp $c_i[\textrm{energy}]$) designates the actual action duration $\textrm{d}_i$ (resp energy $\textrm{e}_i$) it takes to run action $x_i$. Note that the value of $c_i$ is not known a priori; it is observed while performing $x_i$. We have the following estimates of worst cases for action $x_i$:

\begin{itemize}
\item $C_i(\LO)[r]$ represents the optimistic worst-case cost of an action $x_i$ in resource $r$ when the system operates in \LO-mode. In this scenario, the robot executes all planned actions normally. If the actual accumulated costs $\sum_{k}c_k[r]$ exceed the optimistic worst-case accumulated costs based on $C_i(\LO)[r]$ (see section \ref{sec:Approach}), the system switches to \HI-mode, indicating an exceptional environment.
\item $C_i(\HI)[r]$ represents the pessimistic worst-case cost of an action $x_i$ in resource $r$ when the system runs in \HI-mode. As already said, a \LO-action must be able to complete even though the system switches to \HI-mode in the meantime. Indeed, it may be impossible to stop the robot in the middle of an action. Thus, a \LO-action may have a cost in \HI-mode where $\forall r, C_i(\HI)[r] \geq C_i(\LO)[r]$.
  
\end{itemize}

The \mcmcts{} process, as any MCTS process, produces sequences of actions that attempt to maximize the objective function $g$ while respecting resource constraints. It also calculates the budget $b_k(\LO)[r]$ (resp $b_k(\HI)[r]$) required in resource $r$ to reach any action $k$ in \LO (resp \HI) mode.
At run-time, when the system detects that such a budget in \LO-mode for any resource $r$ has been exceeded, it changes mode. This process is detailed in \ref{sec:Approach}. 

\section{Problem Statement} 
\label{sec:Problem}

We aim to produce a robust plan that maximizes the objective function $g$ involving actions for which the costs of their resources are uncertain. The design of such systems traditionally relies on heuristics such as \textit{Monte Carlo Tree Search} (MCTS) to produce a plan. However, these heuristics are poorly adapted to uncertain resource costs, and in particular may fail at runtime to achieve critical objectives. Conversely, while the \textit{Mixed Criticality} (MC) approach defines real-time schedules adaptable to various worst-case execution times contingent upon the probabilities of fault occurrences, it has yet to be adapted to many resources, and to resources other than execution time. The problem is therefore to evolve these two approaches in order to study the benefits to be derived from their synergies.

To integrate the concept of MC in the MCTS heuristic, we need to identify how the costs of actions are considered during the plan building process. Among the four phases of MCTS (see \ref{sec:mcts}), the selection and simulation phases are the ones impacted by the cost uncertainty.
These phases have to be enriched in order to produce a better adapted result, whether the system is operating or transitioning between action sequences of different worst-case estimates.

The transition must also be seamless, and can take place at any point during the execution process, ensuring that future critical actions and those in progress have sufficient resources available.
Additionally, if there are no uncertain costs and if there are only critical actions, our solution should produce results similar to those produced by an MCTS heuristic.

\subsection*{\textbf{Research Objectives}}

Our first objective (RO1) is to develop a solution to the mission planning problem that maximizes the number of \HI-actions completed even in exceptional execution environments where actual resource consumption corresponds to pessimistic assumptions. The resource budgets must be strictly enforced.

Our second objective (RO2) is to maximize the number of \LO-actions completed in normal execution environments where resource consumption corresponds to optimistic assumptions.

\section{Approach}
\label{sec:Approach}

In this section, we present our \mcmcts{} heuristic, an extension of MCTS heuristics to deal with uncertain constraints, such as the energy costs.
Our solution involves running the MCTS heuristic while incorporating the mixed criticality approach during the selection and simulation phases.

\subsection{MCTS phases description}
\label{sec:mcts}

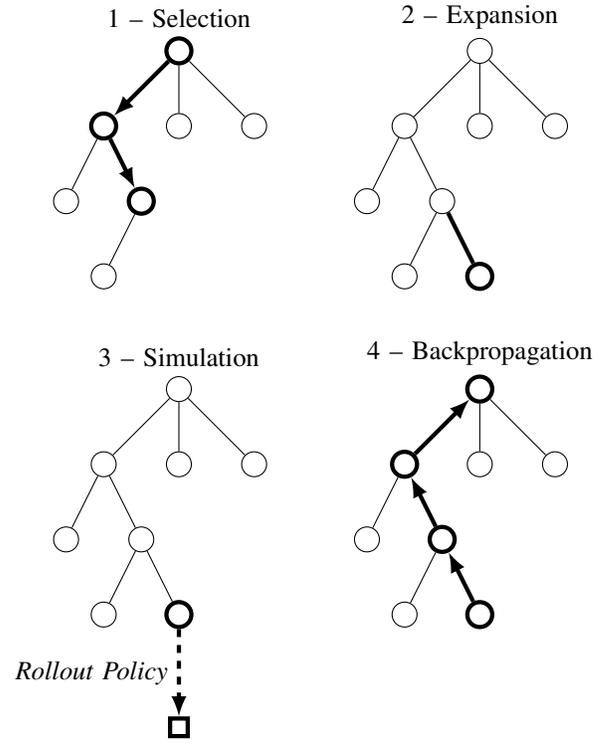
\begin{figure}[htbp!]
\centering
\begin{tikzpicture}
  \begin{scope}[shift={(0,0)}]
    \coordinate[label=above:1 -- Selection, circle, selected, draw] (root) at (3,0);
    \coordinate[circle, selected, draw] (l1) at (2,-1);
    \coordinate[circle, draw] (c1) at (3, -1);
    \coordinate[circle, draw] (r1) at (4, -1);
    \coordinate[circle, draw] (l2) at (1.5, -2);
    \coordinate[circle, selected, draw] (r2) at (2.5, -2);
    \coordinate[circle, draw] (l3) at (2, -3);
    \draw[selected, ->] (root) -- (l1);
    \draw (root) -- (c1);
    \draw (root) -- (r1);
    \draw (l1) -- (l2);
    \draw[selected, ->] (l1) -- (r2);
    \draw (r2) -- (l3);
  \end{scope}

  \begin{scope}[shift={(4,0)}]
    \coordinate[label=above:2 -- Expansion, circle, draw] (root) at (3,0);
    \coordinate[circle, draw] (l1) at (2,-1);
    \coordinate[circle, draw] (c1) at (3, -1);
    \coordinate[circle, draw] (r1) at (4, -1);
    \coordinate[circle, draw] (l2) at (1.5, -2);
    \coordinate[circle, draw] (r2) at (2.5, -2);
    \coordinate[circle, draw] (l3) at (2, -3);
    \coordinate[circle, selected, draw] (new) at (3, -3);
    \draw (root) -- (l1);
    \draw (root) -- (c1);
    \draw (root) -- (r1);
    \draw (l1) -- (l2);
    \draw (l1) -- (r2);
    \draw (r2) -- (l3);
    \draw[selected] (r2) -- (new);
  \end{scope}

  \begin{scope}[shift={(0, -4.5)}]
    \coordinate[label=above:3 -- Simulation, circle, draw] (root) at (3,0);
    \coordinate[circle, draw] (l1) at (2,-1);
    \coordinate[circle, draw] (c1) at (3, -1);
    \coordinate[circle, draw] (r1) at (4, -1);
    \coordinate[circle, draw] (l2) at (1.5, -2);
    \coordinate[circle, draw] (r2) at (2.5, -2);
    \coordinate[circle, selected, draw] (new) at (3, -3);
    \coordinate[circle, draw] (l3) at (2, -3);
    \coordinate[rectangle, selected, draw] (rollout) at (3, -4.5);
    \draw (root) -- (l1);
    \draw (root) -- (c1);
    \draw (root) -- (r1);
    \draw (l1) -- (l2);
    \draw (l1) -- (r2);
    \draw (r2) -- (l3);
    \draw (r2) -- (new);
    \draw[selected, ->, dashed] (new) -- (rollout) node[midway, anchor=east]{\it Rollout Policy};
  \end{scope}

  \begin{scope}[shift={(4, -4.5)}]
    \coordinate[label=above:4 -- Backpropagation, circle, selected, draw] (root) at (3,0);
    \coordinate[circle, selected, draw] (l1) at (2,-1);
    \coordinate[circle, draw] (c1) at (3, -1);
    \coordinate[circle, draw] (r1) at (4, -1);
    \coordinate[circle, draw] (l2) at (1.5, -2);
    \coordinate[circle, selected, draw] (r2) at (2.5, -2);
    \coordinate[circle, draw] (l3) at (2, -3);
    \coordinate[circle, selected, draw] (new) at (3, -3);
    \draw[selected, <-] (root) -- (l1);
    \draw (root) -- (c1);
    \draw (root) -- (r1);
    \draw (l1) -- (l2);
    \draw[selected, <-] (l1) -- (r2);
    \draw (r2) -- (l3);
    \draw[selected, <-] (r2) -- (new);
  \end{scope}
\end{tikzpicture}
\caption{MCTS phases}
\label{fig:MCTS}
\end{figure}

MCTS is composed of four phases: \emph{selection, expansion, simulation} and \emph{backpropagation} \cite{Survey_Browne_2012}. These four phases are executed iteratively to incrementally grow a tree, as shown in \figurename~\ref{fig:MCTS}. Each node of the tree represents a sequence of actions and contains data about the expected reward of the subsequent sequences of actions. During each iteration, a new leaf node is added to the tree, and the statistics within the ancestor nodes are updated accordingly. This process repeats until a \textit{computation budget} is reached. Note that a computation budget is a standard notation in MCTS literature to designate processing limits to which the heuristic expands the tree. It should not be misunderstood with mission (or action) time budgets or worst-case execution time.

During the selection phase, an expandable node of the tree is selected. An expandable node is defined as a node that has at least one child that has not yet been visited during the search. The heuristic begins at the root node of the tree and recursively traverses child nodes until an expandable node is reached. In order to check whether an action $x_i$ is feasible at node $k$, we verify conditions such as whether the accumulated cost after executing $x_i$ does not surpass the budget. Therefore, cost values affect the result of this phase.

During the expansion phase, a leaf node is added to the selected node $k$ by choosing an action $x_{i+1}$ among the possible actions to execute after the sequence $\boldsymbol{x}_i = (x_1, x_2,...,x_i)$. The list of possible actions is impacted by the uncertainty of costs, but was already computed in the selection phase.

In the simulation phase, the expected value of the reward of the new node is computed by simulating possible scenarios after the execution of $\boldsymbol{x}_{i+1}$ using a \textit{rollout policy}. This policy can be a random policy or a heuristic tailored to the problem. A maximum horizon is defined to limit the length of the action sequences simulated within this phase. During the execution of the rollout, it is important to know which sequences of actions are possible after $\boldsymbol{x}_{i+1}$, considering the environment and the budget. This means uncertain costs also make the results of this phase uncertain.

In the backpropagation phase, the result of the simulation phase for the new node is added to the statistics of all nodes along its path up to the root of the tree.
These statistics are usually unbiased estimators of the rollout evaluations for the objective function.
In this phase, the expected utility of each of the ancestor nodes is updated (backpropagated) with a more precise value, as we have more data about the child nodes resulting from the simulation. This phase is not impacted by the uncertainty of costs.

MCTS is an algorithm that can be greatly optimized by rebuilding the plan periodically during the mission execution, a process known as replanning. As the system state is updated, the new plan is better adapted to the new possible outcomes of the mission execution. Replanning thus leads to a more efficient use of the remaining resources by adapting to the new reality of the situation. Thus, our approach can also benefit from replanning, since a change of mode is no substitute for updating the plan to reflect the current system state.

\subsection{\mcmcts{}: MCTS adaptation to uncertain costs}

In our approach, each action has two costs, one in \HI-mode, one in \LO-mode. Thus, each node $k$ in the MCTS action tree has two associated accumulated costs $b_k(\text{\HI})$ and $b_k(\text{\LO})$, with the accumulated costs for the root node $b_0(\text{\HI})$ and $b_0(\text{\LO})$ being zero. The computation of these accumulated costs depend on the action $x_i$ assigned to node $k$.

For \LO-actions, the accumulated costs are calculated as
\begin{equation*}
  \forall r, \left\lbrace
    \begin{aligned}
      b_k(\LO)[r] & = b_{k-\text{1}}(\LO)[r] + C_i(\LO)[r] & (1) &\\
      b_k(\HI)[r] & = b_{k-\text{1}}(\LO)[r] + C_i(\HI)[r] & (2)
    \end{aligned}
  \right.
\end{equation*}
whereas for \HI-actions, the accumulated costs are
\begin{equation*}
  \forall r, \left\lbrace
    \begin{aligned}
      b_k(\text{\LO})[r] & = b_{k-\text{1}}(\text{\LO})[r] + C_i(\text{\LO})[r] &  (3) &\\
      b_k(\text{\HI})[r] & = \max_{h \leq j < k}\left(b_{j}(\text{\HI})[r]\right) + C_i(\text{\HI})[r] & (4)
    \end{aligned}
    \right.
  \end{equation*}
where $h$ is either the node with the last \HI-action in the tree branch of node $k$ or the root node.

In \LO-mode, as all actions are executed, the accumulated costs for \HI and \LO actions are computed the same way. This means the accumulated cost is simply the sum of $C_i(\LO)[r]$ of every node in the branch (equations (1) and (3)).

In \HI-mode, the accumulated cost of a \HI-action $x_i$ must consider the worst outcome of several situations: either the system only ran in \HI-mode during the execution of the current action $x_i$; or it was already running in \HI-mode while executing the previous \HI-action $x_h$; or it switched from \LO to \HI-mode during the execution of one of the previous \LO-actions $x_l$ with $h < l < i$. When node $i$ is added to the tree, its accumulated cost is computed by considering the maximum accumulated cost of these different situations (equation 4). When a budget overrun occurs, the system may change mode during the execution of a \LO-action and must execute the next \HI-action if one exists. However, the cost of executing the \HI-action cannot be precomputed from an unknown intermediary system state where the current action is still being executed. Indeed, the system state is only known at the end of each action. We therefore ensure the current action completes by allocating it a budget in \HI-mode even for a \LO-action. In \HI-mode, as we consider that a \LO-action must complete its execution in \HI-mode during a mode change, its cumulative cost in \HI-mode is computed by adding the cost of executing $x_i$ in \HI-mode to the cumulative cost of executing the previous action in \LO-mode (equation (2)).

Note that in the case where there are no uncertainties, \textit{i.e.} $\forall r, \forall i, C_i(LO)[r]=C_i(HI)[r]=C_i[r]$, then $\forall r, \forall k, b_k(LO)[r]=b_k(HI)[r]=\sum_iC_i[r]$. This means that \mcmcts behaves exactly as MCTS when the action costs do not change between \LO and \HI mode assumptions.

\begin{figure}[htbp!]
  \centering
  \begin{tikzpicture}
    \begin{scope}[xshift=1cm]
      \coordinate[circle, selected, draw] (root) at (3,0);
      \coordinate[circle, selected, draw] (l1) at (2,-1);
      \coordinate[circle, draw] (c1) at (3, -1);
      \coordinate[circle, draw] (r1) at (4, -1);
      \coordinate[circle, draw] (l2) at (1.5, -2);
      \coordinate[circle, selected, draw] (r2) at (2.5, -2);
      \coordinate[circle, draw] (l3) at (2, -3);
      \coordinate[circle, selected, draw] (r3) at (3, -3);
      \coordinate[circle, selected, draw] (new) at (3.5, -4);
      \draw[selected, ->] (root) -- node[midway,above,sloped] { $x_0$ } node[midway,below,sloped] { \footnotesize\LO } (l1);
      \draw (root) -- (c1);
      \draw (root) -- (r1);
      \draw (l1) -- (l2);
      \draw[selected, ->] (l1) -- node[midway,above,sloped] { $x_1$ } node[midway,below,sloped] { \footnotesize\LO } (r2);
      \draw (r2) -- (l3);
      \draw[selected, ->] (r2) -- node[midway,above,sloped] { $x_2$ } node[midway,below,sloped] { \footnotesize\LO } (r3);
      \draw[selected, ->] (r3) -- node[midway,above,sloped] { $x_3$ } node[midway,below,sloped] { \footnotesize\HI } (new);
    \end{scope}
    \begin{scope}[xshift=1cm, yshift=-5cm,xscale=1.6]
      \draw [-latex] (0, 0) -- (4.5, 0) node [right] {$<r>$};
      \foreach [count=\i] \n/\p in {0/\LO,1/\LO,2/\LO,3/\HI}
        \path (\i-1,0) -- node [above] { $x_\n \in \xbs{\p}$ } ++(1,0) --
          +(0,.2) [draw] -- +(0,-.2) node[below] { $b_\n(\LO)[r]$ };
        \end{scope}
    \begin{scope}[yshift=-7.4cm, xshift=1cm, xscale=1.6]
      \draw [-latex] (0, 0) -- (4.5, 0) node [right] {$<r>$};
      \draw (1,0) -- node [above] {$x_1 \in \xbs{\LO}$} (2.2,0) -- +(0,.2) [draw] -- +(0,-.2) node[below]{$b_1(\HI)[r]$};
      \draw(2,-.2) [draw, dashed, ->] -- (2,.7) node[above]{mode change};
      \draw (2.2,0) -- node [above] {$x_3 \in \xbs{\HI}$} (4,0) -- +(0,.2) [draw] -- +(0,-.2) node[below]{$b_3(\HI)[r]$};
      \foreach [count=\i] \n/\p in {0/\LO}
      \path (\i-1,0) -- node [above] { $x_\n \in \xbs{\p}$ } ++(1,0) --
      +(0,.2) [draw] -- +(0,-.2) node[below]{$b_\n(\LO)[r]$};
    \end{scope}
  \end{tikzpicture}
  \caption{\mcmcts{} to scheduler instructions and constraints}
  \label{fig:mcmcts-to-scheduling}
\end{figure}
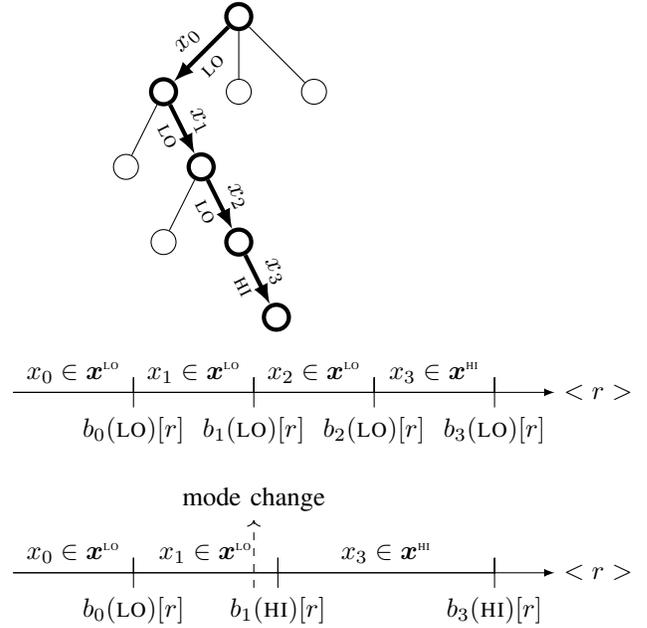

MCTS or \mcmcts{} cannot guarantee that all \HI-objectives will be achieved. In particular, resources may not allow this. However, we want to prevent \LO-objectives from being met at the expense of \HI-objectives. To achieve this, the rewards for \HI-objectives are always set to be greater than the sum of the rewards for \LO-objectives:$\forall x_i\in\boldsymbol{x}^{\HI}, \sum_{x_j \in\boldsymbol{x}^{\LO}} r_j < r_{i}$. The sum of the rewards $r_i$ for accomplishing the objectives are therefore the greatest contributors for the objective function.
  
\figurename~\ref{fig:mcmcts-to-scheduling} illustrates how the outcome from the \mcmcts{} step is mapped into instructions for the scheduler. In this example, the selected action sequence is $(x_0, x_1, x_2, x_3)$. $x_3$ is a \HI-action, while $x_0,x_1, x_2$ are \LO-actions.

For every action $x_k$, the accumulated cost in \LO mode $b_k(\LO)[r]$ from the \mcmcts{} tree is used as the budget corresponding to resource $r$. The systems starts in \LO-mode. After executing any action $x_k$, the consumption of every resource $r$ (\textit{e.g.,}, duration, or energy) since the system has started is compared against $b_k(\LO)[r]$. If the budget is exceeded for at least one resource, the system switches (or stays) in \HI-mode.

When the system is running in \HI-mode, it discards the next action $x_k$ if it is a \LO-action. This ensures that the resources consumption stays below the computed \HI-mode accumulated cost $b_k(\HI)[r]$. In the setup shown on \figurename~\ref{fig:mcmcts-to-scheduling}, only $x_0, x_1, x_2$ can be dropped as $x_3$ is a \HI-action.

\figurename~\ref{fig:mcmcts-to-scheduling} represents an example of a mode change. The first line shows the consumption of resource $r$ in \LO-mode, while the second one is a possible scenario where an action surpasses its \LO-mode cost. Line $<r>$ represents the progressive usage of the available resources. When the mode change occurs, we let \LO-action $x_1$ complete its execution in \HI-mode, \LO-action $x_2$ is discarded and \HI-action $x_3$ is executed after $x_1$ completion.

After transitioning to \HI-mode, it is possible that the system encounters a more favorable environment, resulting in the actual accumulated cost for every resource $r$ returning to a value below $b_k(\LO)[r]$ after executing action $x_k$. In this case, the system reverts to \LO-mode and runs upcoming \LO-actions normally. However, the system needs to check whether any previously dropped \LO-action $x_i$ is a dependency of the \LO-action $x_k$ it is about to execute. If this is the case, then action $x_k$ must be dropped as well. When a replanning occurs, the resources are reallocated completely, and the system restarts in \LO-mode after a new plan has been adopted.

If $b_k(\HI)[r]$ is exceeded for any resource $r$ at the end of $x_k$ execution or at any point before that, either the assumptions used when designing the system were wrong, or the system operates by accident outside its safe operating conditions, for example because of a faulty hardware component. In both those cases, \mcmcts{} still performs more safely than MCTS by dropping \LO-mode actions and giving the system a chance to recover and return within its nominal operating conditions.

\section{Evaluation}
\label{sec:Eval}

In this section, we evaluate the performance of our heuristic on a data collecting scenario with regard to the research objectives listed in section~\ref{sec:Problem}.

\begin{itemize}
\item How much does \mcmcts{} improve the number of completed \HI-actions in exceptional environments while guaranteeing resource constraints (RO1) ?
\item How much does \mcmcts{} improve the number of completed \LO-actions in normal environments while guaranteeing resource constraint (RO2) ?
\end{itemize}

Our benchmark heuristics are two traditional MCTS implementations using two different strategies that each try to accomplish one of these objectives.

\textit{Maximize actions under pessimistic assumptions} (Section~\ref{sec:EvalRep}): we compare our solution to an MCTS implementation that makes pessimistic assumptions about costs during the plan building. Indeed, every \LO or \HI-action in the plan can be executed during the mission even when an exceptional environment occurs (RO1).

\textit{Maximize actions under optimistic assumptions} (Section~\ref{sec:EvalOptRep}): we compare our solution to an MCTS implementation that makes optimistic assumptions about costs, maximizing the number of actions included in the plan (RO2). If an action exceeds its optimistic budget, the MCTS implementation performs a replanning operation after the action has been completed. This additional replanning and all subsequent periodic ones will use pessimistic assumptions.

\subsection{Problem setup}
\label{setup}

Consider a farm scenario where a drone needs to collect data from sensors spread throughout a field. In this scenario, the robot operates within an energy budget and a flight time budget. Some sensors are almost out of battery, and retrieving their data before they run out of energy is highly critical. The energy and flight time necessary for movement are subject to uncertainty, as potential obstacles or environmental conditions such as strong winds may affect the robot's motor efficiency.

The costs in time and energy of an action have a minimum value under optimal conditions. But the more unlikely the conditions, the higher the costs. Consequently, the energy cost and the flight time to move the robot can be reasonably modeled as a half-normal probabilistic distribution. The minimum value is half the worst-case cost in \LO-mode ($C(\text{\LO})$). The standard deviation depends on whether we wish to test the execution on a normal or exceptional environment. In a normal environment, we fix it at $C(\text{\LO})/10$. Otherwise, we fix it at $C(\text{\LO})/3$, increasing the chance of higher costs.

\subsection{Experiment setup}  
We model the terrain as a 100×100 grid. The robot can move freely inside the field. Table~\ref{tab:acCosts} contains the considered cost for moving and for retrieving data. In \HI-mode, each \HI-action is allowed to use twice the previously allocated budget, and \LO-actions are dropped. The energy costs are given in percentage of battery level. The maximum budget $B[energy]$ is fixed to 60\% of battery level in order to evaluate the influence of a shortage of a second resource in the results. We use a random rollout policy and the Upper Confidence Bound for Trees (UCT) \cite{kocsis2006} selection policy, a best-first policy which generates an upper confidence bound to assess the optimality of the selected node.

In our implementation, the actions considered are reaching an objective point, and retrieving the data from the sensor located at that point. The \textit{computation budget} used is 600 and it refers to the number of times MCTS executes the selection phase. The \textit{horizon} used is 5 and represents the maximum length of actions sequences tested in the rollout policy. The \textit{C parameter} in UCT determines how much we prioritize exploration of different paths over exploitation of the best current path. We fix it to 0.5. The distance between the previous objective point and the next one is used to calculate the cost of an action.

The robot starts at a corner and is expected to finish at the opposite corner. Replanning is made every 2 actions so as to allow the robot running \mcmcts{} to execute a part of the mission in \HI-mode, as replanning after every action would bring the robot back to \LO-mode immediately after every mode change. Fifty different scenarios have been generated, each featuring 15 randomly positioned targets with 4 of them being of high criticality. We run MCTS and \mcmcts{} 100 times on each random scenario and evaluate the results with different time budgets.

The objective function considered is
\begin{equation}
  \label{eq:objfun}
  g(\boldsymbol{x}, t) = \frac{\sum_{i \in \boldsymbol{x}}(r_i) - \frac{t}{B[time]}*10^{-4}}{7.5}
\end{equation}
where $\boldsymbol{x}$ is the set of actions already completed, $r_i$ is the reward for completing action $x_i$, $t$ is the time consumed from the beginning of the mission until the end of the last action and $B[time]$ is the total time budget available.

The $\frac{t}{B[time]}$ expression is used to prioritize plans that achieve the same objectives in less time. The $10^{-4}$ weight is used to ensure this value is always below the reward of achieving an objective. The $7.5$ weight ensures the total value is always below $1$, which is necessary for UCT. For \mcmcts{}, we use $b(LO)[time]$ as $t$, as it is the accumulated cost for the normal mode of operation. The reward for reaching the recharge site is 1.0, while the one for performing any other HI- action is 0.2 and for completing any \LO-action 0.0166. Note that $\forall x_i\in\boldsymbol{x}^{\HI}, \sum_{x_j \in\boldsymbol{x}^{\LO}} r_j < r_{i}$ as \HI-actions are our priority. As reaching the recharge site is the main \HI-action we want to ensure is in the plan, its reward is greater than the sum of the rewards of the other \HI-actions.

We want to assess each algorithm's ability to find a plan where the robot retrieves data from objective points and reaches the recharging site before exhausting its allocated budget. A robot can fail to reach the recharging site due to unexpected high action costs. In this situation, the number of achieved objectives will be counted as zero.

\begin{table}[htb]
    \caption{Action costs}
    \centering
    \begin{tabular}{c|c|c|c|c}
    Action & C\textsubscript{i}(\LO) & C\textsubscript{i}(\LO) & C\textsubscript{i}(\HI)& C\textsubscript{i}(\HI)\\
    & [duration] & [energy] & [duration] & [energy] \\
    \hline
    Move (one unit)  & 2.0   & 0.2\% & 4.0 & 0.1\% \\
    Retrieve data    & 5.0  & 1.0\% & 10.0  & 2.0\% \\
    \end{tabular}
    \label{tab:acCosts}
\end{table}

\subsection{MCTS plans built on pessimistic costs}
\label{sec:EvalRep}
We evaluate the performance of \mcmcts{} when compared to MCTS when the latter is configured to only generate plans that may never require more resources than the budget given.

When considering (RO1) where we only consider the exceptional situation, this MCTS is optimal as it is designed to never surpass the allocated budget. Indeed, every action in the plan will always be executed. In the most pessimistic scenario, \mcmcts{} will drop every \LO-action and execute a plan that only contains \HI-actions. It will compute the accumulated cost as the sum of $C(HI)[r]$, just as the pessimistic MCTS configuration. Therefore, in an exceptional environment, both approaches will execute similar amounts of \HI-actions, making them equal when it comes to (RO1).

When considering a normal environment (RO2), MCTS may execute less \LO-actions than \mcmcts{} due to its pessimism. Thus, we evaluate the number of objectives achieved by \mcmcts{} compared with that of MCTS by simulating situations where the budget in \LO mode is never exceeded.

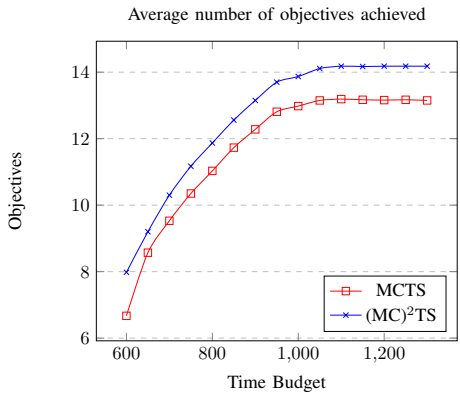
\begin{figure}[htbp!]
\centering
\begin{tikzpicture}[scale=0.7]  
  \begin{axis}[
    title={Average number of objectives achieved},
    xlabel={Time Budget},
    ylabel={Objectives},
    yticklabel = {\hspace{1.8em}\pgfmathprintnumber{\tick}},
    ymajorgrids=true,
    grid style=dashed,
    legend pos=south east,
    ]
    \addplot[mark=square,smooth,red] table [y index=1] {objectives_replanning.txt};
    \addlegendentry{MCTS}
    \addplot[mark=x,smooth,blue] table [y index=2] {objectives_replanning.txt};
    \addlegendentry{(MC)$^2$TS}
  \end{axis}
\end{tikzpicture}
\caption{MCTS plans built on pessimistic costs executed in normal environment}
\label{fig:replanning}
\end{figure}

The experimental results are shown in \figurename~\ref{fig:replanning}. The values reach a peak due to the limited energy budget. \mcmcts{} outperforms MCTS in this scenario when the budget is not big enough to accomplish all objectives. This is due to \mcmcts{} having an accumulated cost closer to the execution in a normal environment, allowing it to explore more action sequences thanks to the previously spared budget.

\subsection{MCTS plans built on optimistic costs}
\label{sec:EvalOptRep}
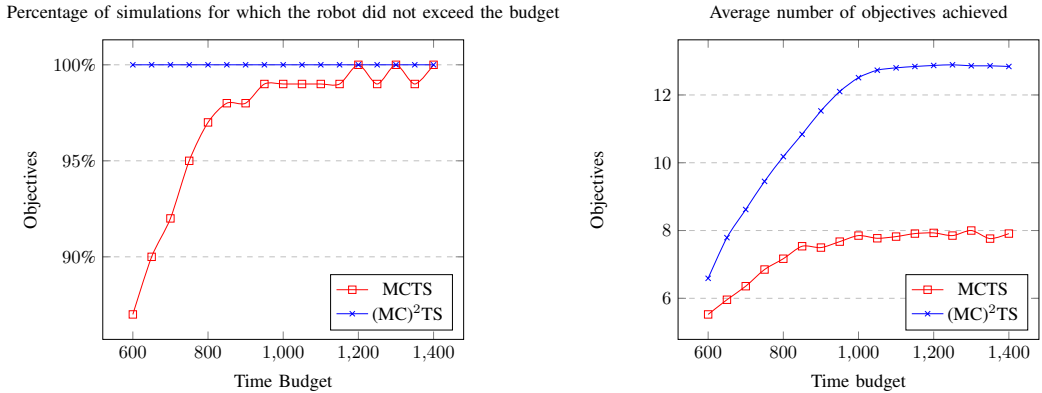
\begin{figure*}[htbp!]
\centering
\subfloat{
\begin{tikzpicture}[scale=0.7]
  \begin{axis}[
    title={Percentage of simulations for which the robot did not exceed the budget},
    xlabel={Time Budget},
    ylabel={Objectives},
    yticklabel ={\pgfmathparse{\tick*100}\pgfmathprintnumber{\pgfmathresult}\%},
    ymajorgrids=true,
    grid style=dashed,
    legend pos=south east,
    ]
    \pgfplotstableread{failed_samples_replanning.txt}\results
    \pgfplotstablecreatecol[
    create col/assign/.code={
      \def\row{\pgfplotstablerow}
      \getthisrow{1}\fail
      \pgfmathsetmacro{\result}{1 - \fail} 
      \pgfkeyslet{/pgfplots/table/create col/next content}\result
    }
    ]{inversed}\results
    \addplot[mark=square,smooth,red] table [y=inversed] {\results};
    \addlegendentry{MCTS}

    \pgfplotstablecreatecol[
    create col/assign/.code={
      \def\row{\pgfplotstablerow}
      \getthisrow{2}\fail
      \pgfmathsetmacro{\result}{1 - \fail} 
      \pgfkeyslet{/pgfplots/table/create col/next content}\result
    }
    ]{inversedMC}\results
    \addplot[mark=x,smooth,blue] table [y=inversedMC] {\results};
    \addlegendentry{(MC)$^2$TS}
  \end{axis}
\end{tikzpicture}
}
\subfloat{
\begin{tikzpicture}[scale=0.7]
  \begin{axis}[
    title={Average number of objectives achieved},
    xlabel={Time budget},
    ylabel={Objectives},
    yticklabel = {\hspace{1.8em}\pgfmathprintnumber{\tick}},
    ymajorgrids=true,
    grid style=dashed,
    legend pos=south east,
    ]

    \pgfplotstableread{objectives_opt_replanning.txt}\results
    \pgfplotstableread{failed_samples_replanning.txt}\weights

    \pgfplotstablecreatecol[
    create col/assign/.code={
      \def\row{\pgfplotstablerow}
      \getthisrow{1}\obj
      \pgfplotstablegetelem{\row}{[index] 1}\of{\weights} 
      \pgfmathsetmacro{\result}{\obj * (1 - \pgfplotsretval)} 
      \pgfkeyslet{/pgfplots/table/create col/next content}\result
    }
    ]{weighted}\results

    \addplot[mark=square,smooth,red] table [y=weighted] {\results};
    \addlegendentry{MCTS}

    \pgfplotstablecreatecol[
    create col/assign/.code={
      \def\row{\pgfplotstablerow}
      \getthisrow{2}\obj
      \pgfplotstablegetelem{\row}{[index] 2}\of{\weights} 
      \pgfmathsetmacro{\result}{\obj * (1 - \pgfplotsretval)} 
      \pgfkeyslet{/pgfplots/table/create col/next content}\result
    }
    ]{weightedmc}\results

    \addplot[mark=x,smooth,blue] table [y=weightedmc] {\results};
    \addlegendentry{(MC)$^2$TS}
    
  \end{axis}
\end{tikzpicture}
}
\caption{MCTS plans built on optimistic costs executed in exceptional environment}
\label{fig:optreplanning}
\end{figure*}

We evaluate the performance of \mcmcts{} when compared to MCTS in case the latter is configured to generate plans that optimize the number of actions in a normal environment. As the plans do not guarantee resource constraints in the most pessimistic scenarios, these plans may fail to ensure a safe return to the recharging site.

When considering (RO2) where we try to optimize the normal case, this MCTS configuration is designed to allow plans with a greater number of actions. Therefore, if we compare its performance to \mcmcts{} in situations where the budget in \LO-mode is never exceeded, it will often achieve more objectives by not respecting the resource constraints.
However, when considering an exceptional environment (RO1), MCTS will generate plans that require more resources than the allocated ones to complete. Therefore, we evaluate how often MCTS plans fail to handle worst case situations by simulating it in exceptional environments. We also evaluate the number of total objectives \mcmcts{} accomplishes when compared to MCTS during these missions.

The experimental results are shown in \figurename~\ref{fig:optreplanning}. The first graph shows that with lower budgets optimistic MCTS fails midway through the execution of the mission multiple times, reducing its effectiveness. However, as shown in the second graph, \mcmcts{} is able to accomplish more objectives even with higher budgets. This is due to its ability to return to \LO-mode once cumulative costs have been reduced to values below the optimistic assumptions.

\subsection{Computational budget influence}
An essential consideration is whether the results from previous simulations remain consistent across different computational budgets, and how this value influences the comparison between the heuristics. To evaluate this behavior, we simulate the situation with pessimistic plans using a time budget of 600 and varying the computational budget.

\begin{figure}[htbp!]
  \centering
  \begin{tikzpicture}[scale=0.7]
    \begin{semilogxaxis}[
      title={Average number of objectives achieved},
      xlabel={Computational Budget},
      ylabel={Objectives},
      yticklabel = {\hspace{1.8em}\pgfmathprintnumber{\tick}},
      ymajorgrids=true,
      grid style=dashed,
      legend pos=north west,
      ]
      \addplot[mark=square,smooth,red] table [y index=1] {objectives_comp.txt};
      \addlegendentry{MCTS}
      \addplot[mark=x,smooth,blue] table [y index=2] {objectives_comp.txt};
      \addlegendentry{(MC)$^2$TS}
    \end{semilogxaxis}
  \end{tikzpicture}
  \caption{Simulation varying the computational budget in a normal environment}
  \label{fig:compbud}
\end{figure}
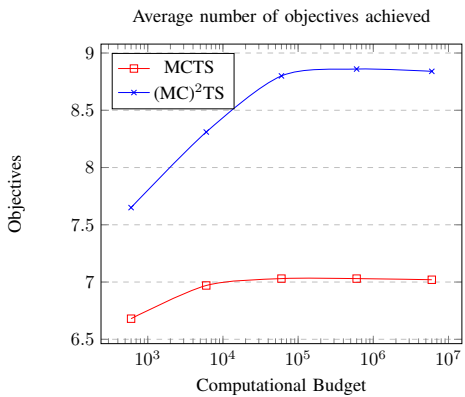

The experimental results are shown in \figurename~\ref{fig:compbud}. As the computational budget increases, the algorithms are capable of finding more optimal plans. With higher budgets, both algorithms degenerate into a standard tree search, and they reach a peak number of objectives. As \mcmcts{} cost assumptions are more optimistic, it consistently achieves more objectives on average than pessimistic MCTS.

\subsection{Conclusions}
We compared \mcmcts{} to two extreme MCTS approaches: one that prioritizes (RO1), and one that prioritizes (RO2). The strategy with pessimistic assumptions may ensure the plan can be executed, but it underperforms when evaluating the number of actions executed. Similarly, the strategy with optimistic assumptions optimizes the actions executed, but fails to provide a plan that can be executed even in the worst case. Therefore, \mcmcts{} is a better alternative if both objectives are desirable, as it a reaches a good compromise between ensuring budget constraints and maximizing actions.

\section{Related Works}
\label{sec:RelatedWork}
In this section, we explore the literature on MCS and MCTS, as these two domains play a key role in our proposed framework for optimizing robot operations.

\subsection{Mixed Criticality Systems}
\label{sec:MC}

In Mixed-Criticality Systems, researchers have focused on Real-Time Scheduling \cite{Burns2017}. As we consider Decision Making Heuristics, we rather focus on the different system models than on the MC scheduling algorithms themselves.

Two mechanisms are usually considered to address a resource failure event : either discard the \LO-tasks in \HI-mode, or degrade the quality of \LO-tasks \cite{Burns2013}. Gu et al. have criticized the strategy of discarding \LO-tasks in \HI-mode \cite{Arvind2019} and propose an unused budget reclamation scheme. Though resource-efficient to some extent, it adopts a pessimistic outlook, degrading the overall efficiency potential.
A few studies have proposed to attempt the minimum service level of \LO-tasks in \HI-mode. Liu et al. propose continuing to execute \LO-tasks in \HI-mode but with a smaller WCET \cite{Liu2016}. However, such an alternative is outside the scope of our case study. Several works also propose increasing the tasks' period in \HI-mode \cite{Huang2015, Zhu13, Lipari98}, but our system is not periodic.

In the context of energy-aware mixed-criticality systems, some studies propose Dynamic Voltage and Frequency Scaling (DVFS) or DVFS with Earliest Deadline First (EDF-VD) scheduling \cite{Liu2016, Huang2014}. While these methods aim to reduce energy consumption by gracefully degrading \LO-tasks in \HI-mode, they lack inherent prioritization of \HI-tasks over \LO-tasks. In these studies, DVFS serves merely a mechanism for degrading the execution of a \LO-task. Our approach differs in that it considers energy as an uncertain parameter of the problem, and not just as a resource to be managed. Therefore, we treat energy (as well as action duration) as a first-class citizen within the MC system, equally important as execution time.

\subsection{Monte-Carlo Tree Search}
\label{sec:MCTS}

Monte-Carlo Tree Search (MCTS) has proven to be pivotal in addressing complex decision-making problems, such as gaming, planning, optimization, and scheduling \cite{Hofmann2022, Survey_Browne_2012}.
Its exploration-exploitation properties make it particularly valuable in path planning applications. Dam et al. extended the application of Monte Carlo methods to improve exploration strategies for robot path planning \cite{Dam2022}.
Kartal et al. proposed hybrid approach, combining MCTS with Branch and Bound for multi-robot action allocation problem with time windows and capacity constraints \cite{Kartal2016}. These constraints and time windows only consider one lower and upper bound, which differs from the MC approach.

Most existing studies consider a known environment where cost is fixed. Given a dynamic environment, Patten et al. have proposed a method for using different samples of the robot's current knowledge to simulate future events \cite{patten2018monte}. However, it does not consider the uncertainties associated with the robot's actions. Unlike other approaches that may consider the uncertainty of the rewards of an action \cite{Survey_Browne_2012, Maciej2021_survey, Castellini2020, Li2023}, we specifically focus on the uncertainties associated with the robot's resources, introducing a novel aspect of uncertain costs. 

Additionally, we highlight mode-change and replanning in \mcmcts{}. We demonstrate how they are complementary techniques that increase the safety and efficiency. This emphasizes our adaptability even in the face of future cost changes.

\section{Conclusions and Perspectives}
\label{sec:Conclusion}
In this article, we tackle the challenge of action planning amid uncertain action costs. Estimating these costs in complex systems can be challenging and often results in overestimations. Consequently, this leads to oversized systems, resulting in the waste of resources and reduced performance.

We reformulate the Mixed Criticality (MC) approach, originally proposed in the real-time community, in two ways. Firstly, we generalize this approach to several different resources, including energy and action duration, whereas it was originally dedicated solely to execution time. Secondly, we apply this approach to Monte Carlo Tree Search instead of Real-Time Scheduling.

We propose \mcmcts{} an extension of MCTS to mixed-criticality systems. First, we adapt the action planing system model to the MC approach. Next, we extend the cost evaluation to match the different criticality modes of the MC system.

We evaluate our proposal in relation to our initial research objectives, and demonstrate the considerable advantages of our approach in reducing oversizing of the system architecture in the presence of uncertain costs. On an active perception example, we demonstrate that the robot can anticipate changes in the environment and therefore, changes in costs. This anticipation prevents the robot from being lost. Furthermore, the objectives are effectively met since the costs are not as consistently pessimistic as they would be with a traditional MCTS approach.

As part of our future work, we will extend our model to a swarm of drones, and to enrich distributed robot planning with the Mixed-Criticality approach. We will also design a middleware to demonstrate the applicability of our approach in a real environment.

\section*{Acknowledgment}
This material is based upon work supported by the CIEDS (Interdisciplinary Centre for Defence and Security of Institut Polytechnique de Paris) and by the FARO project (Heuristic Foundations of Robot Swarms).

\printbibliography
\end{document}